\documentclass[aps,twocolumn,preprintnumbers,amsmath,amssymb,amsfonts,groupedaddress,superscriptaddress,prl,showpacs,showkeys,floatfix]{revtex4-1}

\usepackage{graphicx}
\usepackage{subfigure}
\usepackage{bm}
\usepackage{amsmath}
\usepackage{color}
\usepackage{array}
\usepackage{CJK}

\newcommand{\bmath}{\begin{mathletters}}
\newcommand{\emath}{\end{mathletters}}
\newcommand{\be}{\begin{eqnarray}}
\newcommand{\ee}{\end{eqnarray}}
\newcommand{\ba}{\begin{array}}
\newcommand{\ea}{\end{array}}
\newcommand{\la}{\langle}

\newcommand{\ra}{\rangle}

\newcommand{\pr}{\prime}

\begin{document}
\title{Experimental realization of a fast controlled-Z gate via a shortcut-to-adiabaticity }

\author{Tenghui Wang}
 \affiliation{Zhejiang Province Key Laboratory of Quantum Technology and Device, Department of Physics, Zhejiang University, Hangzhou, 310027, China}
\author{Zhenxing Zhang}
 \affiliation{Zhejiang Province Key Laboratory of Quantum Technology and Device, Department of Physics, Zhejiang University, Hangzhou, 310027, China}
 \author{Liang Xiang}
 \affiliation{Zhejiang Province Key Laboratory of Quantum Technology and Device, Department of Physics, Zhejiang University, Hangzhou, 310027, China}
 \author{Zhilong Jia}
 \affiliation{Key Laboratory of Quantum Information, University of Science and Technology of China, Hefei, 230026, China}
 \author{Peng Duan}
 \affiliation{Key Laboratory of Quantum Information, University of Science and Technology of China, Hefei, 230026, China}
 \author{Zhiwen Zong}
 \affiliation{Zhejiang Province Key Laboratory of Quantum Technology and Device, Department of Physics, Zhejiang University, Hangzhou, 310027, China}
 \author{Zhenhai Sun}
 \affiliation{Zhejiang Province Key Laboratory of Quantum Technology and Device, Department of Physics, Zhejiang University, Hangzhou, 310027, China}
 \author{Zhangjingzi Dong}
 \affiliation{Zhejiang Province Key Laboratory of Quantum Technology and Device, Department of Physics, Zhejiang University, Hangzhou, 310027, China}
 \author{Jianlan Wu }
 \affiliation{Zhejiang Province Key Laboratory of Quantum Technology and Device, Department of Physics, Zhejiang University, Hangzhou, 310027, China}
 \author{Yi Yin}
 \email{yiyin@zju.edu.cn}
 \affiliation{Zhejiang Province Key Laboratory of Quantum Technology and Device, Department of Physics, Zhejiang University, Hangzhou, 310027, China}
 \author{Guoping Guo}
 \email{gpguo@ustc.edu.cn}
 \affiliation{Key Laboratory of Quantum Information, University of Science and Technology of China, Hefei, 230026, China}

\begin{abstract}

For a frequency-tunable two-qubit system, a controlled-Z (CZ) gate can be realized by adiabatically driving the qubit system
through an avoided level crossing between an auxiliary state and computational levels.
Here, we theoretically propose a fast CZ gate using a shortcut-to-adiabaticity (STA).
Experimentally, the STA CZ gate is implemented with a 52 ns control pulse for two coupled superconducting Xmon qubits.
Measured fidelity of the STA CZ gate is higher than 96.0\%, in both quantum process tomography and randomized benchmarking.
The protocol allows a flexible design of the evolution time and control waveform. We suggest that
this `fast adiabatic' CZ gate can be directly applied to other multi-qubit quantum systems.
\end{abstract}

\maketitle


\section{I. Introduction}

Quantum logic gates are the building blocks of quantum circuits in quantum computation~\cite{ChuangBook,LaddNat10}.
The gate-based quantum computation requires a combination of single-qubit gates and a two-qubit
entangling gate. Especially, the two-qubit gate is the foremost element in complex
quantum algorithms. A controlled-Z (CZ) gate is a relatively common two qubit
entangling gates, from which a controlled-NOT gate can also be generated~\cite{WendinRep17}.
Among different physical quantum systems, Superconducting qubit system has been one of the most promising
candidates for quantum computation~\cite{LaddNat10,WendinRep17}.
Due to the lack of a `ZZ' coupling~\cite{DiCarloNat09,GhoshPRA13,MartinisPRA14,BarendsNat14}
in superconducting transmon or Xmon qubits, it is difficult to realize the CZ gate directly in computational levels
of each qubit (the ground $|0\rangle$ and excited $|1\rangle$ states). Instead, a CZ gate
has been proposed to use non-computational energy levels (the second excited $|2\rangle$ state)~\cite{StrauchPRL03}.
Driven adiabatically near the avoided level crossing between the two-qubit state $|11\ra$ and $|20\ra$,
the $|11\rangle$ state can acquire a state-dependent phase, with other computational
states unchanged. If the phase is designed as $\pi$, a controlled $\pi$-phase gate or the CZ gate can be realized.

However, a long time is required in the adiabatic evolution. Inevitable errors are introduced within
the long procedure, due to the qubit decoherence and non-adiabatic leakage.
In order to suppress such errors, Martinis \textit{et al.} propose and realize a `fast adiabatic' protocol of CZ gate, where a
fast and specially designed drive reduces non-adiabatic errors as much as possible~\cite{MartinisPRA14,BarendsNat14}.
Another `fast adiabatic' method is to use a shortcut-to-adiabaticity (STA) to completely eliminate non-adiabatic transitions~\cite{DemirplakJPCA03,BerryJPhysA09,XChenPRL2010,MasudaPRS10,Torrontegui13AAMOPhy,CampoPRL12,CampoPRL13,TongSR15,SantosSR15}.
By introducing a counter-dibatic (CD) field, the STA protocol can force the quantum state
to remain in the instantaneous eigenstate of the reference Hamiltonian. The STA protocol has
been extensively applied to control the state evolution of a single qubit, in cold atoms~\cite{BasonNatPhy11,DuYXNatCommu16},
NV centers~\cite{JFZhangPRL13,ZhouNatPhys16}, trapped-ion~\cite{AnNatCommu16} and superconducting
qubits~\cite{ZZXPRA17,SCPMA2018,WangNJP2018,ZZXNJP2018}. Applying the STA protocol in a multi-qubit system is yet a nontrivial task.

In this paper, we propose and implement a CZ gate for two coupled superconducting Xmon qubits, using a STA protocol.
In the subspace of $|11\rangle$ and $|20\rangle$ states, the STA protocol is directly applied to derive a `fast adiabatic'
waveform. A practical problem is that this drive requires a variable complex coupling between qubits, instead of
a fixed coupling in the real system. The problem is resolved by introducing a representation transformation and a
rescaling method. Following the theoretical model, the protocol is experimentally realized in our Xmon qubit system.
The measurement fidelity are over 96\% for both the quantum process tomography (QPT) and the randomized benchmark (RB).
An interleaved RB is also implemented with a fidelity about 94\%. The fidelity can be further improved, with
a better multi-qubit system in the future.

\section{II. Theoretical Protocol}
For two coupled Xmon qubits, the system Hamiltonian is
\be
\label{}
H_{\mathrm{sys}} = H_{\mathrm{qA}}+H_\mathrm{qB}+H_\mathrm{c},
\ee
where $H_\mathrm{qA(B)}$ is the single qubit Hamiltonian, and $H_\mathrm{c}$ is the coupling term.
With lowest three levels of a Xmon considered, the single qubit Hamiltonian $H_\mathrm{qA(B)}$ can be expressed as
\be
\label{}
H_\mathrm{qA(B)} = \hbar \omega_\mathrm{qA(B)} |1\ra\la 1| + (2\hbar\omega_\mathrm{qA(B)}+\hbar\Delta) |2\ra\la 2|   ,
\ee
where $\hbar$ is the reduced Planck constant, $\hbar\omega_\mathrm{qA(B)}$ is the energy difference between the
ground $|0\rangle$ and excited $|1\rangle$ states of the qubit Q$_\mathrm{A(B)}$, and $\Delta=\omega^{12}_\mathrm{q}-\omega^{01}_\mathrm{q}$
is the anharmonicity for both Xmons. The coupling term $H_\mathrm{c}$ is expressed as
\be
\label{}
H_c = \hbar g(J_\mathrm{A}^\dagger\cdot J_\mathrm{B} + \mathrm{H.c.}),
\ee
where $g$ is the coupling strength, $J = |0\ra\la1| + \sqrt{2} |1\ra\la2|$ is the lowering
operator for a three level system, and H.c. stands for the Hermitian conjugate.

\subsection{A. Adiabatic evolution in subspace}
We consider a subspace including two-qubit states $|11\rangle$ and $|20\rangle$. In both states, the
1st and 2nd numbers represent the qubit state of Q$_\mathrm{A}$ and Q$_\mathrm{B}$, respectively. The energy of $|11\rangle$
is $\hbar(\omega_\mathrm{qA}+\omega_\mathrm{qB})$, while the energy of $|20\rangle$ is $\hbar(2\omega_\mathrm{qA}+\Delta)$.
In our system, the difference between two qubit frequencies is initially set at
$(\omega_\mathrm{qA}-\omega_\mathrm{qB})/2\pi\approx 550$ MHz.  With the anharmonicity $\Delta/2\pi\approx-240$ MHz,
the energy of $|20\rangle$ is initially much larger than that of $|11\rangle$.

If we fix $\omega_\mathrm{qB}$ and slowly lower $\omega_\mathrm{qA}$, the energy difference between $|20\rangle$ and
$|11\rangle$ will decrease correspondingly. As shown in the energy level diagram (Fig.~\ref{fig1}(a)),
this detuning will gradually bring $|11\rangle$ into an avoided level crossing with $|20\rangle$.
Following a designed adiabatic trajectory (black dashed line in Fig.~\ref{fig1}(a)), an initialized
$|11\rangle$ state will adiabatically evolve along the instantaneous eigenstate (solid blue line) and
be brought back to the $|11\rangle$ state. Without a non-adiabatic leakage, a dynamic phase
$\varphi$ is accumulated in the final state $|\psi\rangle = e^{i\varphi}|11\rangle$. The dynamic phase
$\varphi=-\int(\omega_\mathrm{qA}(t)+\omega_\mathrm{qB})dt+\varphi^\prime$, including an extra phase $\varphi^\prime$
in addition to the trivial dynamic phase of single Xmons. Because the subspace Hamiltonian can be written as
\be
\label{}
H_{\mathrm{sub}}(t) = \hbar \begin{bmatrix}
\omega_\mathrm{qA}(t)+\omega_\mathrm{qB}& \sqrt{2}g \\
\sqrt{2}g & 2\omega_\mathrm{qA}(t)+\Delta
\end{bmatrix},
\ee
we obtain the extra phase $\varphi^\prime = \int{\sqrt{2}g\tan(\frac{\theta(t)}{2})dt}$, where
$\theta$ is the polar angle with $\tan\theta(t)=2\sqrt{2}g/\Delta_d(t)$, and
$\Delta_d(t)=\omega_\mathrm{qA}(t)-\omega_\mathrm{qB}+\Delta$ is the frequency difference between $|11\rangle$ and $|20\rangle$.

For a state initialized at $|00\rangle$, $|01\rangle$, or $|10\rangle$, the same
detuning procedure will not induce extra dynamic phase because
these states are off-resonance with both avoided-level crossings. Then the total unitary
operator for the computational level can be written as
\be
\label{}
U = |00\ra\la 00| + e^{-i\int \omega_\mathrm{qA}(t)dt}|10\rangle\langle10| \notag \\+ e^{-i\int \omega_\mathrm{qB}dt}|01\rangle\langle01| + e^{i\varphi}|11\rangle\langle11|.
\ee
Removing the trivial dynamic phase of single Xmons, the unitary operator reduces to
\be
\label{}
U = |00\ra\la 00| + |10\rangle\langle10| + |01\rangle\langle01| + e^{i\varphi^\prime}|11\rangle\langle11|,
\ee
where $\varphi^\prime=\int{\sqrt{2}g\tan(\frac{\theta(t)}{2})dt}$ is the control phase.
By adjusting the evolution path of $\theta(t)$, we can implement an arbitrary two-qubit control phase gate.

\subsection{B. STA protocol and rescaled Hamiltonian}
In general, a time-dependent Hamiltonian $H_0(t)$ can be expanded in its instantaneous eigen basis,
i.e., $H_0(t)=\sum_n\epsilon_n(t) |\psi_n(t)\rangle\langle \psi_n(t)|$, where
$\epsilon_n(t)$ is the $n$-th eigenenergy and $|\psi_n(t)\rangle$ is the $n$-th eigenstate.
According to the STA protocol, a counter-diabatic Hamiltonian $H_{\mathrm{cd}}(t)$ is formally
written as~\cite{BerryJPhysA09}
\be
H_{\mathrm{cd}}(t) = i\hbar\sum_n \big[|\partial_t \psi_n(t)\rangle\langle \psi_n(t)| \notag \\-\langle \psi_n(t)| \partial_t \psi_n(t)\rangle | \psi_n(t) \rangle \langle \psi_n(t)| \big].
\label{eq_07}
\ee
When the system is driven with a total Hamiltonian of $H_0(t)+H_\mathrm{cd}(t)$,
the non-adiabatic transition can be suppressed within a short operation time.
In the subspace of $|11\rangle$ and $|20\rangle$, we apply the STA protocol to the
previous adiabatic trajectory. The counter-diabatic Hamiltonian is calculated as
$H_{\mathrm{cd}}(t)=-\dfrac{1}{2}\hbar\dot{\theta}(t)\{-i|11\ra\la20|+i|20\ra\la11|$\}.
Practically, we can not physically generate an imaginary coupling term in the subspace.
To remove the imaginary coupling, we introduce an unitary transformation
$U_{\mathrm{rot}}(t)=|11\ra\la11|+e^{-i\phi(t)}|20\ra\la20|$. After shifting the diagonal energy,
the subspace Hamiltonian is rewritten as~\cite{IbanezPRL12}
\be
\label{}
H_{\mathrm{sub}}^\prime(t) = U_{\mathrm{rot}}(t)\left[H_{\mathrm{sub}}(t)+H_{\mathrm{cd}}(t)\right]U_{\mathrm{rot}}^\dagger(t) \notag\\+i\dot{U}_{\mathrm{rot}}(t)U_{\mathrm{rot}}^\dagger(t)= \hbar \begin{bmatrix}
0& \Omega(t)\\
\Omega(t) & \Delta_d(t)+\dot{\phi}(t)
\end{bmatrix},
\ee
where $\Omega(t)=\sqrt{2g^2+\dot{\theta}(t)^2/4}$, and $\phi(t)$ is the azimuth angle
in the Hamiltonian of $H_{\mathrm{sub}}(t)+H_{\mathrm{cd}}(t)$ with $\tan(\phi(t))=-\dot{\theta}(t)/2\sqrt{2}g$.
In $H_{\mathrm{sub}}^\prime(t)$, the off-diagnoal term $\Omega(t)$ is time-dependent,
which can be realized if $g$ is a tunable coupling between two qubits~\cite{ChenPRl14,RoushanNatPhys17,RoushanSci17}.
In our Xmon sample, the capacitive coupling is fixed between neighboring qubits,
resulting in a fixed off-diagonal term of $\sqrt{2}g$ in the subspace Hamiltonian.
Then we need to find a new rescaled Hamiltonian to realize the same function as with $H_{\mathrm{sub}}^\prime(t)$.

We divide the original $H_{\mathrm{sub}}^\prime(t)$ to $N$ segments, each of duration $\Delta t$.
With a sufficiently large $N$, the unitary operator for the $m$-th duration is
\be
\label{}
U_m = \exp\{-iH_{\mathrm{sub}}^\prime(m\Delta t)\Delta t\}.
\ee
To acquire a constant off-diagonal term, the new rescaled Hamiltonian
$H_{\mathrm{sub;new}}^\prime$ of the $m$-th segment can be defined as
\be
\label{rescalH}
H_{\mathrm{sub;new}}^\prime(\Delta \tau_m)= H_{\mathrm{sub}}^\prime(m\Delta t)\cdot \dfrac{\sqrt{2}g}{\Omega(m \Delta t)},
\ee
and the duration time for the $m$-th segment $H_{\mathrm{sub;new}}^\prime(\Delta \tau_m)$ is rescaled as
\be
\label{rescalt}
\Delta\tau_m= \Delta t \cdot \Omega(m\Delta t)/\sqrt{2}g.
\ee
The unitary operator $U_m$ is kept the same as before for each segment, although the time is rescaled to
fix the off-diagonal term of the subspace Hamiltonian. Correspondingly, the time-dependent form of
$\omega_\mathrm{qA}$ is rewritten as $\omega_\mathrm{qA}(\tau)=\Delta_d(\tau)+\dot{\phi}(\tau)+\omega_\mathrm{qB}-\Delta$.
Combining the unitary transformation~\cite{IbanezPRL12} and the rescaling approach~\cite{JFZhangPRL13}, we apply the STA protocol in
the coupled Xmon qubit system to implement a fast control phase gate.

\section{III. Experimental Setup}

\begin{figure}[htp]
\centering
\includegraphics[width=1.0\columnwidth]{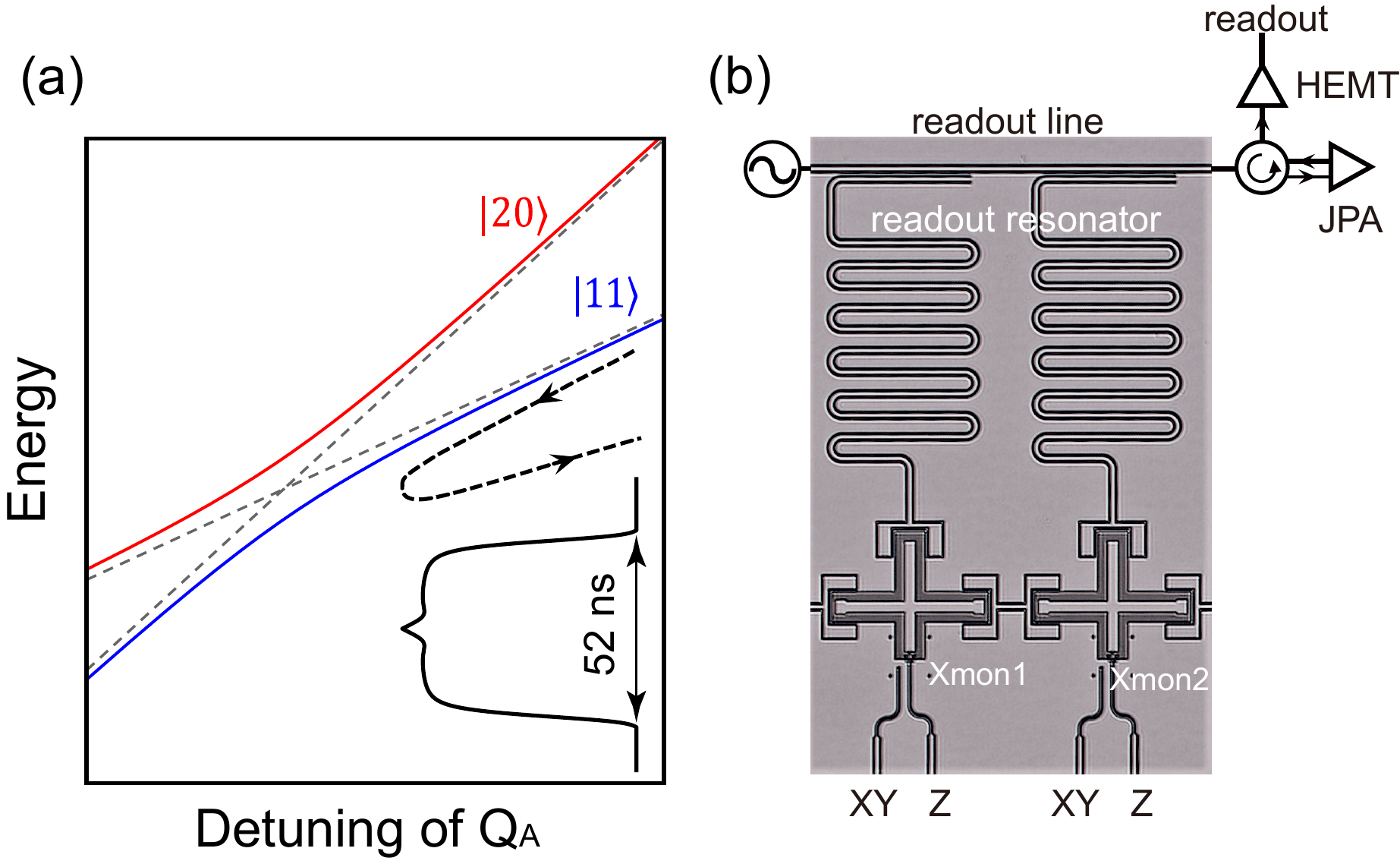}
\caption{(a) An energy level diagram of the avoided level crossing. The dashed line with black arrows
represents an adiabatic trajectory. The solid black line is the CZ gate pulse, proportional to
the actual frequency tuning of Q$_\mathrm{A}$.
(b) An optical micrograph of two coupled Xmon qubits.}
\label{fig1}
\end{figure}

Figure~\ref{fig1}(b) displays an optical micrograph of two coupled Xmon qubits on a chip sample.
The fabrication of the chip is the same as described before~\cite{WangNJP2018}. The qubit chip is mounted in an
aluminum sample box and cooled in a dilution refrigerator whose base temperature is about 10 mK.
For each Xmon, four arms of the cross are connected to a readout resonator (top), control lines (bottom) and
neighboring Xmons. Through the $Z$ control line, a flux current is supplied to bias the Xmon qubit at an operation frequency.
In our experiment, two Xmon qubits are initially biased at $\omega_\mathrm{qA}/2\pi=5.52$ GHz and
$\omega_\mathrm{qB}/2\pi=4.97$ GHz, respectively. The qubit anharmonicity is $\Delta/2\pi\approx-240$ MHz for both Xmons.
At these operation points, the energy relaxation time $T_1$ are 14.4 $\mu$s and 12.9 $\mu$s, and
the pure decoherence time $T_2^*$ are 12.3 $\mu$s and 3.5 $\mu$s for Q$_\mathrm{A}$ and Q$_\mathrm{B}$ respectively. The second qubit is biased
far away from the sweet point, resulting in a relative shorter decoherence time.
The $XY$ control line provides a microwave drive signal to the qubit to
manipulate the qubit state.

At the end of any qubit manipulation, the qubit state is encoded in a coupled
readout resonator, and can be detected by a dispersive readout.
The bare frequency of readout resonators are $\omega_\mathrm{rA}/2\pi=6.56$ GHz and
$\omega_\mathrm{rB}/2\pi=6.71$ GHz for Q$_\mathrm{A}$ and Q$_\mathrm{B}$, respectively.
In the dispersive readout, a microwave measurement signal is sent through the readout line, interacting with the
readout resonator. After amplified by a Josephson parametric amplifier (JPA)~\cite{RoyAPL15} and a high electron
mobility transistor (HEMT), the readout signal is finally collected.
The readout fidelity of the ground state $|0\rangle$ and excited state $|1\rangle$
are $F_\mathrm{qA}^{0}=96.1\%$, $F_\mathrm{qB}^{0}=94.5\%$ and $F_\mathrm{qA}^{1}=88.6\%$, $F_\mathrm{qB}^{1}=86.4\%$ for Q$_\mathrm{A}$ and Q$_\mathrm{B}$ respectively.
In addition, the on chip wire-bonding is applied across the control line to reduce the $Z$ crosstalk.
The measured crosstalk coefficients are -6\%(Q$_\mathrm{B}\rightarrow$Q$_\mathrm{A}$) and 4\%(Q$_\mathrm{A}\rightarrow$Q$_\mathrm{B}$)~\cite{BarendsNat14}.

\section{IV. Results}
\label{sec4}
\begin{figure}[htp]
\centering
\includegraphics[width=1.0\columnwidth]{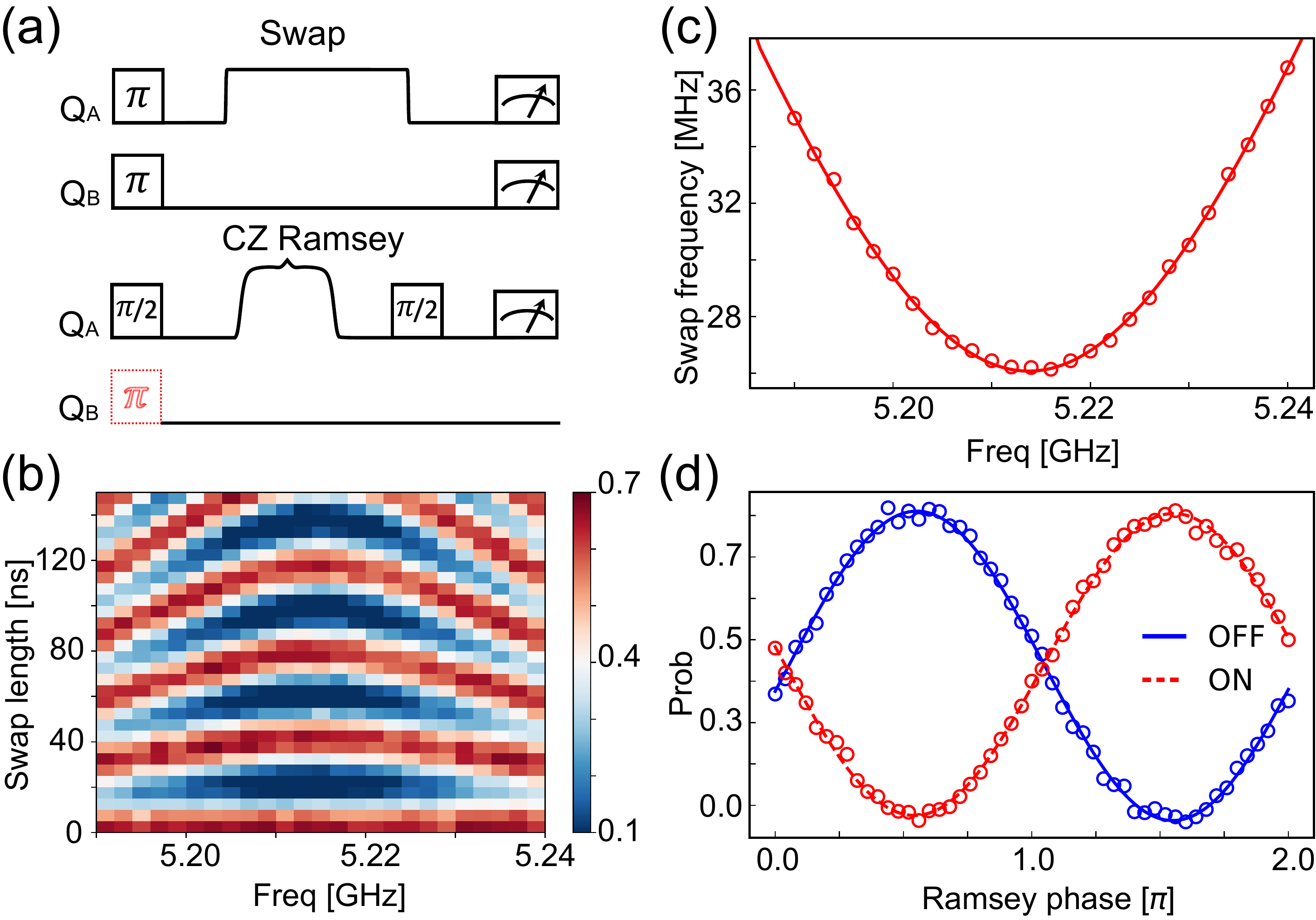}
\caption{(a) The control sequence for the swap spectroscopy and Ramsey fringe experiment (b) The probability
of $|11\ra$ versus the qubit frequency $\omega_\mathrm{qA}$ and the swap length. (c) The extracted
swap frequency $\omega_\mathrm{swap}$ versus qubit frequency $\omega_\mathrm{qA}$, shown in red circle.
The fitting curve is shown in red lines. (d) The probability of Q$_\mathrm{A}$ is displayed as a function of
a Ramsey phase. The blue (red) circle represents the Ramsey experiment results for
Q$_\mathrm{B}$ initialized to $|0\rangle(|1\rangle)$. Solid lines are the corresponding fitting curves.}
\label{fig2}
\end{figure}

To obtain accurate parameters of the coupled two qubits, we measure a swap spectroscopy between
$|11\rangle$ and $|20\rangle$ states. In Fig.~\ref{fig2}(a), we show a schematic pulse sequence of
the swap spectroscopy. By applying a $\pi$-pulse to each qubit, two qubits are initially prepared
in the $|11\rangle$ state. Then a rectangle detuning pulse is applied to Q$_\mathrm{A}$ to lower its
frequency $\omega_\mathrm{qA}$. When $\omega_\mathrm{qA}$ equals $\omega_\mathrm{res}=\omega_\mathrm{qB}-\Delta$,
the state $|11\rangle$ will resonate with $|20\rangle$. The states of two qubits
are finally measured simultaneously, with a dispersive readout. In Fig.~\ref{fig2}(b), we plot the measured
probability $P_{|11\rangle}$ versus the detuning time and the detuned $\omega_\mathrm{qA}$. A typical
chevron pattern can be observed, which reveals a quantum state oscillates between $|11\rangle$ and $|20\rangle$.
Theoretically, the probability $P_{|11\rangle}$ oscillates with a swap frequency
$\omega_\mathrm{swap}=\sqrt{8g^2+(\omega_\mathrm{qA}-\omega_\mathrm{res})^2}$.
For a specific $\omega_\mathrm{qA}$, we take the fourier transform of the $P_{|11\ra}(t)$ oscillation,
from which $\omega_{\mathrm{swap}}$ is extracted. In Fig.~\ref{fig2}(c), we plot the extracted
oscillation frequency as a function of $\omega_\mathrm{qA}$, and make a curve fitting with the theoretical
formula of $\omega_\mathrm{swap}$. The fitting result leads to an accurate estimation of the coupling
strength $g/2\pi\approx9.19$ MHz and the resonant frequency $\omega_\mathrm{res}/2\pi\approx5.214$ GHz.
These two parameters facilitate the following design of a STA waveform for the CZ gate.

In principle, the control pulse can be designed with different functions,
whenever they are compatible with the theoretical protocol.
In this work, we select a Hanning-Window function for $\dot{\theta}(t)$, i.e.,
$\dot{\theta}(t)=\dfrac{\theta_{f}-\theta_{i}}{T}\{1-\cos(2\pi t/T)\}$, with $0<t<T$ for the first
half of the trajectory. The polar angle $\theta_i=\arctan(2\sqrt{2}g/\Delta_{d}(0)) \approx0.047$ is the initial polar
angle in our system, and $\theta_f$ is the maximum polar angle in the trajectory. For the CZ gate,
the control phase $\phi^\prime$ is $\pi$, leading to a maximum polar angle $\theta_f \approx 2.36$.
After $\theta(t)$ reaches the maximum angle $\theta_f$, the 2nd half of the trajectory is applied immediately, as
$\dot{\theta}(t>T)=\dfrac{\theta_{i}-\theta_{f}}{T}\{1-\cos(2\pi (t-T)/T)\}$ ($T<t<2T$). The total time $2T$ is chosen
to be 40 ns in our design. For the whole trajectory, the function of $\omega_\mathrm{qA}$ is given
by $\omega_\mathrm{qA}(t)=2\sqrt{2}g/\tan(\theta(t))+\omega_\mathrm{res}+\dot{\phi}(t)$.
This control function gives $\dot{\phi}(t=0,2T)=0$, which allows a smooth change of $\omega_\mathrm{qA}(t)$
at the start and end of the sequence. Finally, the rescaled $\omega_{q1}(\tau)$ can be obtained with
Eq.~\ref{rescalH} and Eq.~\ref{rescalt}. The total evolution time is about 52 ns after rescaling.
The rescaled pulse for the CZ gate is shown in the inset of Fig.~\ref{fig1}(a), where an obvious
protuberance can be observed in the middle.

Before applying the designed pulse to implement a CZ gate, we need to determine the additional dynamic phase
accumulated in a single qubit, which is measured in the following Ramsey fringe experiment~\cite{YamamotoPRB10}.
A $\pi/2$-pulse is initially applied to Q$_\mathrm{A}$ to create a superposition state of $|0\rangle+|1\rangle$.
A CZ pulse is then applied. The detuning of $\omega_\mathrm{A}$ induces an extra dynamic phase for Q$_\mathrm{A}$.
A second $\pi/2$-pulse is finally applied to complete the Ramsey fringe process.
If there is no CZ pulse in the middle between two $\pi/2$ pulses, Q$_\mathrm{A}$ will be brought to the
first excited state $|1\rangle$. If the second $\pi/2$-pulse is applied with a variable phase (Ramsey phase),
Q$_\mathrm{A}$ state will project to $|1\rangle$ with a cosinusoid probability. The additional dynamic phase
will, however, shift the phase of the cosinusoid function.
Figure~\ref{fig2}(d) presents the final probability of $P_{|1\ra}$ versus the Ramsey phase.
We could observe that the maximum $P_{|1\ra}$ appears at a finite phase, instead of the zero phase.
This shifted phase equals the accumulated dynamic phase we need. A cosinusoid fitting gives us an accurate
phase value to compensate the additional dynamical phase during CZ gate.
Furthermore, to confirm the operation of CZ gate, we compare the Ramsey fringe experiment with Q$_\mathrm{B}$
initialized in $|0\rangle$ or $|1\rangle$. If Q$_\mathrm{B}$ is initially excited to $|1\rangle$, Q$_\mathrm{A}$
will acquire an extra controlled $\pi$ phase, compared to the previous Ramsey fringe experiment.
In Fig.~\ref{fig2}(d), we plot the results of two Ramsey fringe experiments, with Q$_\mathrm{B}$
initialized in $|0\rangle$ or $|1\rangle$. A $\pi$ phase difference can be clearly observed, which verifies
the operation of our CZ gate.

To quantify the CZ gate fidelity, we perform a quantum process tomography (QPT) of the CZ gate.
In the QPT procedure, the output state is obtained
through a map of the input state~\cite{ChuangBook}, i.e.,
\be
\text{\large $\varepsilon$}: \rho\mapsto \text{\large $\varepsilon$}(\rho) = \sum_{i=1}^{16} E_i\rho E_i^+,
\label{eq_19}
\ee
where $\rho$ the initial density matrix of the two-qubit system. Each linear operators $E_{i=1, \cdots, 16}$ can be
expanded by a fixed set of operators $\{\tilde{E}_m, m=1, \cdots, 16\}$, giving $E_i = \sum_m e_{im} \tilde{E}_m$.
The operator basis $\tilde{E}_m$ can be acquired from the  Kronecker product of
pauli operators $ \{I,\sigma_x, \sigma_y, \sigma_z\}$ of each qubit.
The output density matrix can then be rewritten as
\be
\text{\large $\varepsilon$}(\rho) = \sum_{mn} \chi_{mn} \tilde{E}_m\rho \tilde{E}_n^+,
\label{eq_20}
\ee
with $\chi_{mn} = \sum_i e_{im}e_{in}^\ast$. The $\chi$ matrix thus completely characterizes the behavior of a specific gate,
although including errors in the state preparation and measurement.

\begin{figure}[htp]
\centering
\includegraphics[width=1.0\columnwidth]{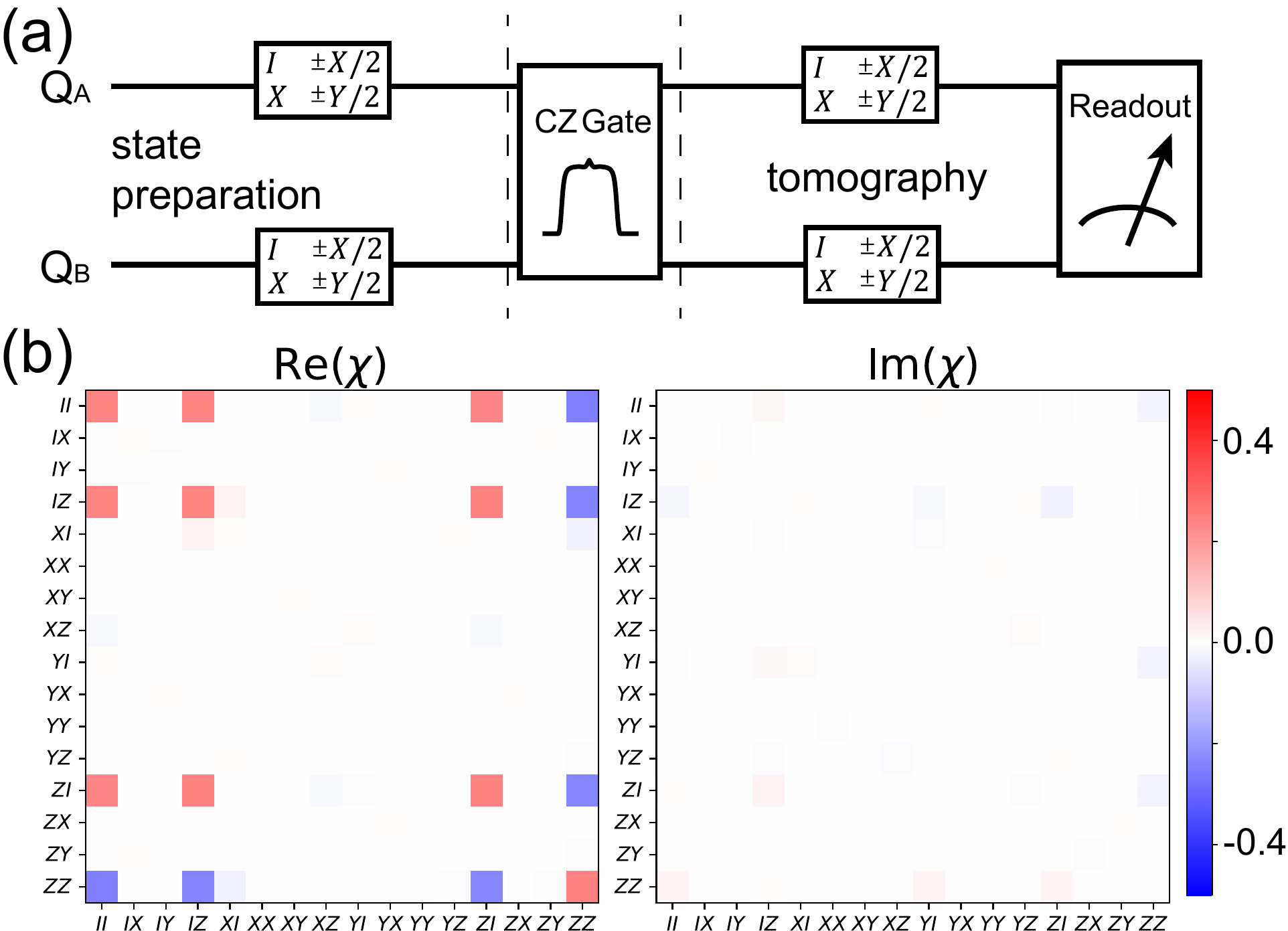}
\caption{(a) The control sequence of the QPT measurement. (b)The experimental measurement of $\chi$ matrix for the CZ gate. }
\label{fig3}
\end{figure}

Figure~\ref{fig3}(a) shows the pulse sequence for the QPT. Different input states are initially prepared,
from the set $\{|0\rangle, |1\rangle, (|0\rangle \pm |1\rangle )/\sqrt{2},(|0\rangle \pm i|1\rangle )/\sqrt{2}\}$ for each qubit~\cite{BialczakNatPhys10,YamamotoPRB10,ChuangBook}.
A CZ pulse is then applied. Afterwards, the output state is measured by the quantum state tomography (QST).
The $\chi$ matrix is numerically calculated by solving Eq.~(\ref{eq_20}).
The experimental result of the $\chi$ matrix is plotted in Figs.~\ref{fig3}(b).
Consistent with the theoretical prediction of a $\chi$ matrix for an ideal
CZ gate, the dominant elements are the operator of $\sigma_z$ and $I$.
To quantify the fidelity of the whole quantum process, we calculate the process fidelity using
$F_\mathrm{P} = \mathrm{Tr}\{\chi \chi_\mathrm{ideal}\}$~\cite{ChuangBook},
with a result of $F_\mathrm{P} = 96.59\%$.
To figure out the error source, we compare our result with a numerical calculation.
Without decoherence, the calculated process fidelity is 99.94\%, which means that
our STA protocol can realize a CZ gate with very high fidelity in the ideal situation.
With decoherence parameters considered, the numerical simulation gives a process fidelity of 98.43\%.
Compared with our experimental result, we could presume that the qubit decoherence is
one of the main loss sources of the process fidelity, together with some other residual control errors.

\label{sec4d}

\begin{figure}[htp]
\centering
\includegraphics[width=1.0\columnwidth]{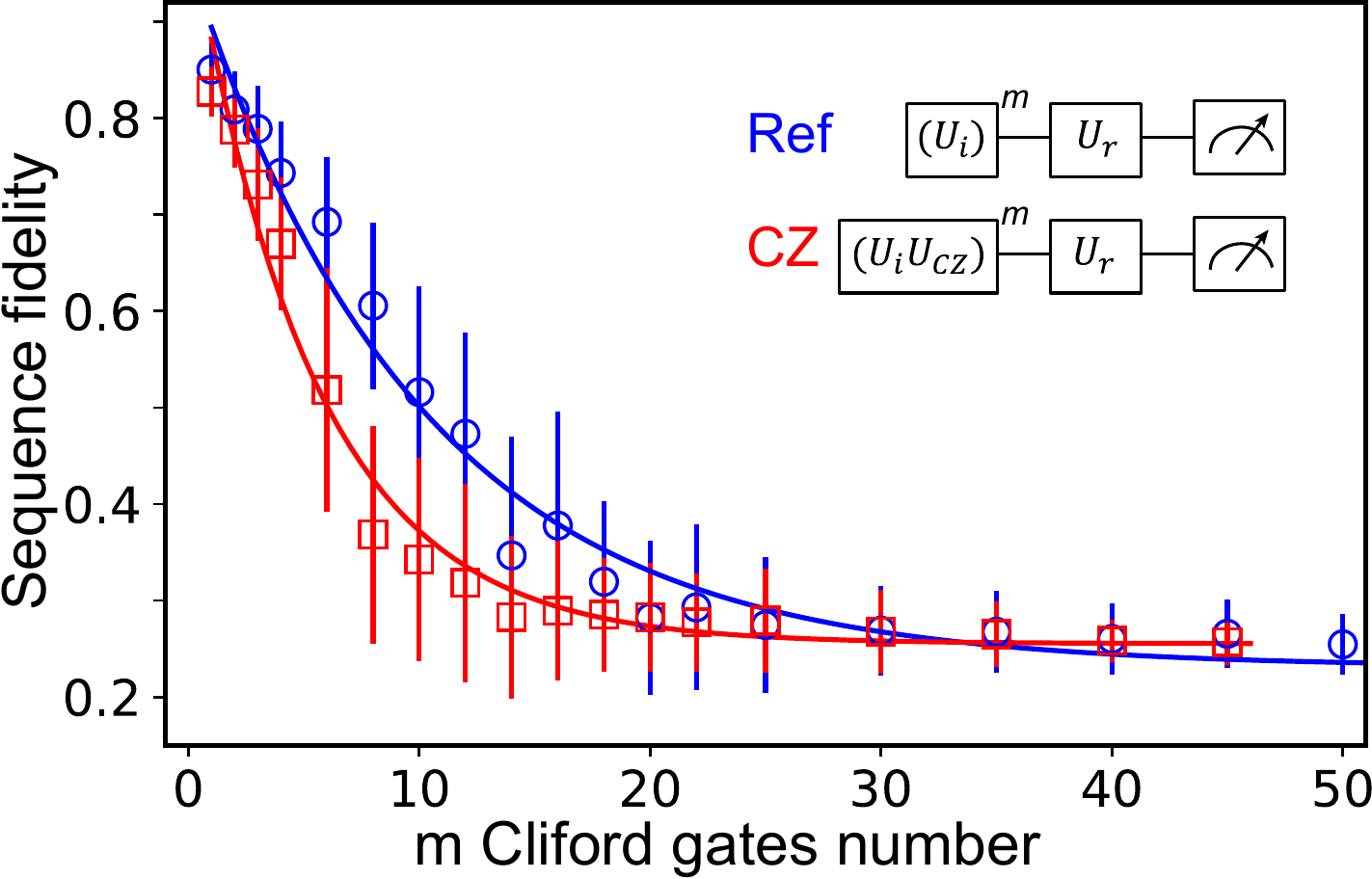}
\caption{The results of Randomized benchmarking measurement. The reference and interleaved
sequence fidelities are displayed as functions of the number of Cliffords. Each sequence fidelity is averaged over $k=40$ randomized
operation. The standard deviation is displayed as an error bar.}
\label{fig4}
\end{figure}

In the above QPT measurement, the errors of state preparation and readout are mixed with the error of a quantum gate operation.
To separately extract the gate fidelity, we perform a Clifford-based RB
measurement~\cite{KnillPRA08,MagesanPRL12_RB,ChowPRL09,BarendsNat14,SheldonPRA16}.
For a two-qubit system, the Clifford group consists of 11520 gate operations.
In principle, each Clifford gate can be realized by a combination from the set of
single qubit gates $\{I, X_{\pi}, X_{\pm \pi/2}, Y_{\pi}, Y_{\pm \pi/2}\}$ and CZ gate.
As shown in the pulse sequence in the inset of Fig.~\ref{fig4}, two qubits are initially
prepared at the $|00\ra$ state, and then driven by a sequence of $m$ randomly
selected Clifford gates. A unitary matrix, $U_C = \prod_{i=1}^m U_i$, describes the combined operation.
$U_C$ still belongs to the closed set of Clifford group. The $(m+1)$-th step reverses
the previous combined operations, and the total quantum operation
can be expressed as $U_\mathrm{tot} = U^\dagger_C \prod_{i=1}^m U_i$.
At the end we measure the remaining population $P_{|00\ra}(t_f)$ of the initial state.
This whole process has been repeated for $k$ (= 40 in our experiment) times, and we
calculate the average result of $P_{|00\ra}(t_f)$ as a function of the Clifford gate numbers $m$.
In Fig.~\ref{fig4}, this sequence fidelity has been fitted by a power-law decaying function~\cite{MagesanPRL12_RB},
$P_{|00\ra}(m) = A_0 p_\mathrm{ref}^m+B_0$, in which $p_\mathrm{ref}$ is a reference depolarizing parameter,
and $A_0$ and $B_0$ include errors in state preparation and readout. With the depolarizing parameter,
the average error over randomized Clifford gates is calculated as $r_\mathrm{ref} = \frac{d-1}{d}(1-p_\mathrm{ref})$,
where $d=2^2=4$ is the Hilbert space dimension for the two-qubit system.
The average error consists of single gates error and CZ gate error, $r_\mathrm{ref} = \dfrac{33}{4}r_\mathrm{SQ}+\dfrac{3}{2}r_\mathrm{CZ}$.
In our experiment, the average error and single qubit error are $r_\mathrm{ref}=0.0712$ and $r_\mathrm{SQ}=0.0017$, respectively.
The average CZ gate fidelity is calculated to be $1-r_\mathrm{CZ}=96.19\%$.

We also make an interleaved operation~\cite{MagesanPRL12_RB} to extract the CZ gate fidelity.
The pulse sequence is also shown in the inset of Fig.~\ref{fig4}, in which
the CZ gate is interleaved in the randomly select Clifford operator.
With the product operator for each step, $U^\pr_C = \prod_{i=1}^m (U_\mathrm{CZ} U_i)$, and the $(m+1)$-th operator of $(U^\pr_C)^\dagger$, we
describe the total operation as $U^\pr_\mathrm{tot}=(U^\pr_C)^\dagger\prod_{i=1}^m (U_\mathrm{CZ} U_i)$~\cite{BarendsNat14,MagesanPRL12_RB}.
The sequence fidelity $P^\pr_{|00\ra}(m)$ is similarly measured. As shown by the red circle in Fig.~\ref{fig4}, $P^\pr_{|00\ra}(m)$
can also be fitted by a power-law decaying function, giving a new depolarizing parameter $p_\mathrm{CZ}$.
Then we calculate the CZ gate fidelity by
\be
F_\mathrm{g}  = 1-\frac{d-1}{d}\left(1-\frac{p_\mathrm{CZ}}{p_\mathrm{ref}}\right).
\label{eq_29}
\ee
In this interleaving RB measurement, the CZ gate fidelity is $F_\mathrm{g}=93.76\%$, which is smaller than the above average fidelity.
This difference may be from the imperfect Z control.

\section{V. Summary}
\label{sec5}
We propose a method to realize a fast CZ gate using the STA protocol.
Through a representation transformation and a rescaled Hamiltonian, we achieve
a `fast adiabatic' evolution with only qubit frequency control. In the absence of
the qubit decoherence, the QPT fidelity of numerical calculation is over 99.9\%,
proven to be a high fidelity CZ gate in the ideal situation.
As an example, we experimentally implement this CZ gate in two coupled superconducting Xmon qubits.
Experimental parameters are acquired from the swap spectroscopy and Ramsey fringe experiment.
From the QPT and RB measurement, the CZ gate fidelities are confirmed to be above 96\%.
An interleaved RB experiment is also performed to give a fidelity of about 94\%, suggesting
a control error from the residue settling in Z pulse. Our protocol provides a feasible `fast adiabatic'
method of CZ gate. In principle, the protocol allows a large flexibility in the evolution
time and control waveform, and can be directly applied in other quantum systems. The fidelity can
be further increased, with the sample quality and control accuracy improved in the future.

\section{acknowledgements}
 \begin{acknowledgements}
The work reported here is supported by the National Basic Research Program of China (2014CB921203, 2015CB921004),
the National Key Research and Development Program of China (2016YFA0301700),
the National Natural Science Foundation of China (NSFC-21573195, 11625419),
the Fundamental Research Funds for the Central Universities in China, and the Anhui Initiative in Quantum Information Technologies (AHY080000).
This work was partially carried out at the University of Science and Technology of China Center for Micro and Nanoscale Research and Fabrication.
\end{acknowledgements}


\end{document}